\documentclass[twocolumn,showpacs,floatfix,prl,superscriptaddress]{revtex4}

\usepackage{float}
\usepackage{graphicx}
\usepackage{bm}
\usepackage{amsmath}
\usepackage{amssymb}
\usepackage{hyperref}
\usepackage{slashed}

\newcommand{\beq}{\begin{eqnarray}}
\newcommand{\eeq}{\end{eqnarray}}
\newcommand{\beqq}{\begin{eqnarray*}}
\newcommand{\eeqq}{\end{eqnarray*}}
\newcommand{\imth}{\hspace{1pt}\mathrm{i}\hspace{1pt}}
\newcommand{\bpm}{\begin{pmatrix}}
\newcommand{\epm}{\end{pmatrix}}

\begin{document}

\title{Topological Bootstrap: Fractionalization From Kondo Coupling}


\author{Timothy H. Hsieh}
\affiliation{
Kavli Institute for Theoretical Physics, University of California, Santa Barbara, California 93106, USA}
\author{Yuan-Ming Lu}
\affiliation{
Department of Physics, The Ohio State University, Columbus, OH 43210, USA}
\author{Andreas W.W. Ludwig}
\affiliation{
Department of Physics, University of California, Santa Barbara, CA 93106, USA}


\begin{abstract}
We propose a route toward realizing fractionalized topological  phases of matter (i.e. with intrinsic topological order) by literally building on un-fractionalized phases.  
Our approach 
employs a Kondo lattice model in which a gapped electronic system of non-interacting fermions
is coupled to 
non-interacting local moments via the exchange interaction. Using
general entanglement-based 
arguments and explicit lattice models, we show that in this way  gapped spin liquids can be induced in the spin system. 
We demonstrate the power of this ``topological bootstrap'' concept
with
two examples: (1) a chiral spin liquid induced by a Chern insulator and (2) a $Z_2$ spin liquid induced by a superconductor. 
In particular, in the latter example, the toric code is realized as an exactly solvable example of topological bootstrap. Our approach can be generalized to all lattices, higher dimensions, and non-abelian topological orders.
\end{abstract}

\maketitle

The discoveries of the integer and fractional quantum Hall effects have, respectively, ballooned during 
the last decade into two broad classes of topological phases, distinguished by whether or not
there are fractional excitations in the bulk.  The integer quantum Hall (IQH) variety, though less exotic in the bulk
in that it does not possess ``intrinsic topological order",
cannot be adiabatically
deformed into a direct product state as long as certain symmetries are preserved.
This class is most generally referred to as ``symmetry protected topological (SPT) phases \cite{chen2013,pollmann2010,wen2009}.  In contrast, the fractional quantum Hall variety exhibits ``intrinsic topological order'' \cite{wen1990}, manifested by fractionalized bulk excitations obeying anyonic statistics, ground state degeneracy on a torus, and in many cases gapless boundary modes as well.

Despite their common origin of quantum Hall systems, these two classes have differed significantly in terms of experimental accessibility. The SPT/IQH variety has been realized in numerous materials and experimental systems.  These include topological insulators protected by time reversal symmetry \cite{hasan2010} and topological crystalline insulators \cite{fu2011} protected by mirror symmetry \cite{hsieh2012}, spanning different material classes and realizable at room temperature. In contrast, intrinsic topological orders have been limited to fractional quantum Hall (FQH) systems \cite{Tsui1982} and quantum spin liquids (QSL) in frustrated magnets \cite{balents2010}. Given the importance of intrinsic topological order in enabling topological quantum computation \cite{nayak2008, kitaev2003}, a new route for realizing intrinsic topological orders is highly desirable.

In this work, we establish a general ``topological bootstrap'' scheme in which an un-fractionalized phase is used to induce a fractionalized phase via Kondo coupling. This powerful machinery allows us to construct a large class of interacting lattice models, whose ground states realize ``projected wavefunctions''\cite{Anderson1987,Gros1989,Wen1999,Wen2002,Lee2006} for intrinsic topological orders. These have been used extensively as variational wavefunctions for topological orders (such as FQH and QSL states), despite the lack of parent microscopic models--a gap filled by the topological bootstrap. To achieve this, we leverage the bulk topological proximity effect introduced in \cite{Hsieh2016}, in which a free-fermion topological phase in system $A$ induces the ``inverse'' topological phase in a proximate system $B$. While the prior work focused on identical free fermion systems $A$ and $B$, the current work involves spin degrees of freedom in system $B$, thus enabling us to generate interacting topological orders in $B$.


\begin{figure}
\centering
\includegraphics[width=3.5in]{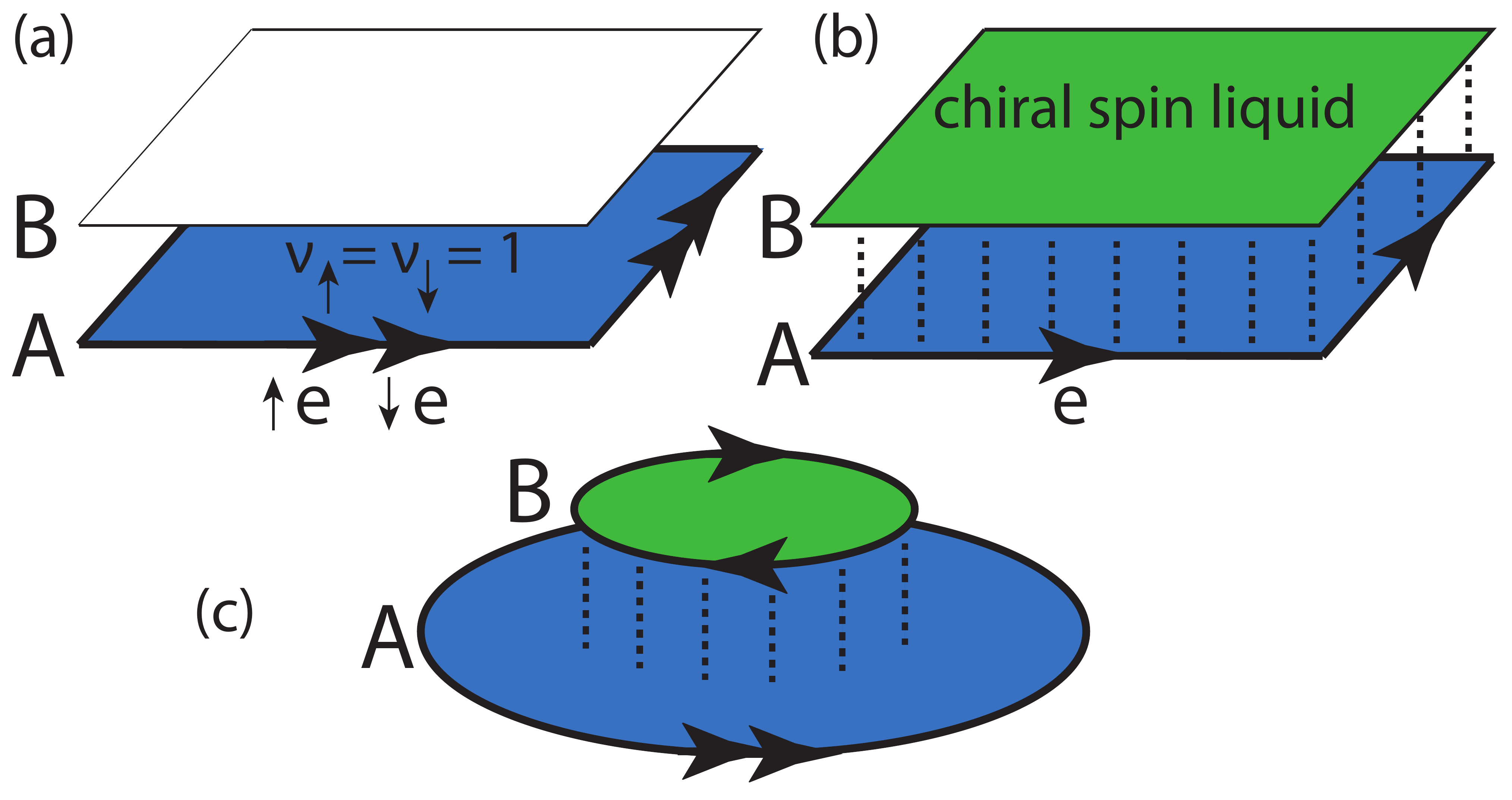}
\caption{Example of topological bootstrap: (a) Two decoupled layers: a Chern insulator (blue, A) and a free layer of spins (white, B). (b) After Kondo coupling, the Chern insulator induces a chiral spin liquid (green) with the opposite chirality.  The counter-propagating spin modes gap out, leaving a charge mode. (c) In a terrace construction, the induced and original edge modes are widely separated and coexist.}
\end{figure}

The topological bootstrap construction has sharp experimental signatures. Our primary example is a Chern insulator for spin up and down electrons, coupled to a free spin system (Fig. 1a).  Before coupling, the Chern insulator has edge modes carrying both spin and charge. The Kondo coupling induces a chiral spin liquid in the spin system, with a spin-carrying edge mode of chirality {\it opposite} to the Chern insulator.  As a result, the counter-propagating spin modes gap out, leaving only a charge mode at the boundary (Fig. 1b).  In contrast, for a terrace construction illustrated in Fig. 1c, the counter-propagating spin mode induced in the spin system is spatially separated from the edge of the electron system and thus can be detected independently.

The structure of this Letter is as follows.  First we will provide general arguments motivating the topological bootstrap. After firmly establishing these arguments in an entanglement-based approach, we will provide a concrete lattice model in which we exactly show that the topological bootstrap realizes a $Z_2$ spin liquid from a free-fermion superconductor. Finally, we present two lattice models in which a chiral spin liquid is induced from a Chern insulator.

\section{General Setup}
\label{LabelSectionGeneralSetUp}

The bulk topological proximity effect considered in \cite{Hsieh2016} involves two free-fermion systems $A$ and $B$ with different Hamiltonians: $H_A$ has a gap $\Delta_A$ and a topologically nontrivial ground state $|\psi^A_0\rangle$, while $H_B=0$ corresponds to a
 system of decoupled fermion degrees of freedom.  The two systems are extensively coupled via
a local  tunneling term, which favors maximal entanglement of  the degrees of freedom  on the corresponding lattice
sites of systems $A$ and $B$ \footnote{See T. Hsieh, Phys. Rev. B 94, 161112(R) (2016) for an entanglement perspective of the bulk topological proximity effect.} .
After coupling, system $B$ perfectly ``screens'' system $A$ by developing the inverse topological phase: for example, a band insulator with Chern number $1$ induces a Chern number $-1$ state in $B$ after the two are coupled.  More precisely, the composite ground state,
as obtained in degenerate perturbation theory  in $g$, the strength of coupling,
 is $|\psi_0\rangle = |\psi^A_0\rangle |\overline{\psi^B_0}\rangle + O(g/\Delta_A)$, where 
the bar denotes the inverse topological phase (such that the composite is topologically trivial).

What if systems $A$ and $B$ have different degrees of freedom, such as free fermions in $A$ and spins in $B$? In this case, it is impossible for $B$ to fully screen $A$ due to the mismatch of Hilbert spaces. However, a partial screening of spin degrees of freedom can induce even more interesting topological phases in system $B$, as will 
become evident below.

Our construction involves composite systems in which $A$ is a 
topological phase of non-interacting fermions and $B$
consists of decoupled spins residing on the same sites as the fermions. In other words, we consider a Kondo lattice model in which the electron system is a topological band insulator or gapped superconductor, as opposed to a metal.  The intuition from the topological proximity effect is that while the spins cannot perfectly screen the fermions by forming the inverse topological phase, they 
are expected to
 inherit the spin sector of the free fermion system. More precisely, in a slave-fermion description of the spins ($\vec{S}=\frac12f^{\dagger} \vec{\sigma} f$), the $f$ fermions 
would  form the inverse phase with respect to $A$, leading to a 
quantum spin liquid  state described by a Gutzwiller-projected wavefunction in system $B$. 
Since the electron system is gapped with zero density of states, we expect that Kondo singlet formation is evaded at weak coupling
($g \ll \Delta_A$), and the above physics is described by a new Kondo mean-field theory \footnote{In progress} with a topologically ordered ground state.

We will illustrate the topological bootstrap in the simplest scenario of a two-dimensional bilayer: $A$ is a 
system of non-interacting fermions at half filling with both spin up and down electrons $c_{\uparrow/\downarrow}$
each  in a Chern number $1$ band, and $B$ is a 
system of decoupled spin-$1/2$s 
with one spin-$1/2$ per site.  The composite Hamiltonian
is
\beq
&\notag\hat H={\hat H}_0+{\hat H}_{K}=\\
&={\hat H}_{\nu=1}(c_{\uparrow}) + {\hat H}_{\nu=1}(c_{\downarrow}) + g \sum_i (c_i^{\dagger} \vec{\sigma} c_i) \cdot \vec{S_i} \label{main}
\eeq
where $H_{K}$ represents 
an antiferromagnetic  Kondo coupling between fermions and localized spins, parametrized by $g \geq 0$. 
We will see that the
 state in the spin system $B$ will be in the same phase as the Gutzwiller projection of the Chern number $1$ state, enforcing one fermion per site.  This is known to be a chiral spin liquid (CSL)\cite{Kalmeyer1987,Wen1989}, 
possessing intrinsic topological order with fractionalized ``semion'' excitations in the bulk, and chiral edge states.

It is useful to directly relate the above Kondo model to the free fermion topological proximity effect. In particular, we can begin with a free-fermion model
\beq
H_{\nu=1}(c_{\uparrow}) + H_{\nu=1}(c_{\downarrow}) + v\sum_{i}\sum_{\sigma=\uparrow,\downarrow}c^\dagger_{i,\sigma}f_{i,\sigma}+~h.c. \nonumber
\eeq
in which the Chern insulator is now coupled to ``Kondo fermions'' $f_{\sigma}$ via a tunneling term of strength $v$.  In \cite{Hsieh2016}, it was found that for any value of $v$, the composite system is always gapped and topologically trivial.  In other words, an $SU(2)$-symmetric Chern insulator of opposite chirality is induced in the system of  ``Kondo fermions"  by an arbitrarily small coupling $v$. Consider a coupling $v$ much smaller than the band gap $\Delta_A$ of the Chern insulator; then
a hopping  term of ``Kondo fermions"
of the order $t\equiv v^2/\Delta_A$
is induced.

Now we add an on-site Hubbard interaction
\beq
U\sum_i(f^{\dagger}_{i,\uparrow}f_{i,\uparrow}+f_{i,\downarrow}^\dagger f_{i,\downarrow}-1)^2
\eeq
for ``Kondo fermions". In the parameter range $t\ll U\ll v\ll\Delta$, the low-energy physics of the ``Kondo fermions" can be described by localized spins $\vec S_i$, and we recover the original Kondo model (\ref{main}) with effective Kondo coupling $g\sim v^2/\sqrt{U\Delta_A}$.  Thus, we expect that the 
state  induced  in the spin system $B$ via 
our Kondo setup is related to the physics of Chern bands with Hubbard interactions, where chiral spin liquids have been found in previous works\cite{Nielsen2013}.

We now support this intuition with both an entanglement argument and microscopic lattice models.

\section{Entanglement Arguments}

The key feature of the antiferromagnetic Kondo coupling term ${\hat H}_K$ in Eq. (\ref{main})
 is that its ground state is a direct product of singlets:

\begin{eqnarray}
&&|\psi_{AB}\rangle \nonumber \\ 
&&=
\prod_i^{sites} 
\left [
{1\over \sqrt{2}}
\sum_{\sigma_i=\pm \frac{1}{2}} sgn(\sigma_i)
c^\dagger_{i,\sigma_i}|0 \rangle_A
\otimes
 |S^z_i=-\sigma_i \rangle_B
\right ] \nonumber
\\  \label{LabelEqMaxEntangled}
&&={1\over \sqrt{2^N}}
\sum_{\{\sigma_i \}} \Big( \prod_i sgn(\sigma_i) \Big)
|\{\sigma_i\}\rangle_A \otimes
|\{-\sigma_i\}\rangle_B
\end{eqnarray}
(where the sum is over all configurations of spins $\sigma_i$  on the lattice sites)
which is a maximally entangled state  between systems $A$ and $B$.  
We will first consider one exact limit of the setup in which the maximal entanglement is most manifest, and then we will interpolate between this limit and the original setup.

We begin with one limit in which we show that the induced spin state in system B is
{\it exactly} the 
Gutzwiller projection of the Chern insulator ground state. 
In this limit, instead
of the {\it local}  Kondo coupling, we consider a coupling which directly projects onto the maximally entangled state $|\psi_{AB}\rangle$
(``{\it global}  projector").  The composite Hamiltonian is thus
\beq
\label{LabelEqGlobalProjector}
H=H_{\nu=1}(c_{\uparrow}) + H_{\nu=1}(c_{\downarrow}) -g |\psi_{AB}\rangle\langle \psi_{AB}|.
\eeq
To
 lowest order in degenerate perturbation theory in the coupling $g$, the effective Hamiltonian for system $B$ is given by
\beq
\langle \psi_B'|H^B_{eff}|\psi_B\rangle = -g\langle \psi^0_A \otimes \psi_B'|\psi_{AB}\rangle\langle \psi_{AB}|\psi^0_A\otimes \psi_B\rangle, \nonumber
\eeq
where $\psi^0_A=|\nu_\uparrow =1\rangle \otimes |\nu_\downarrow =1\rangle$ is the Chern insulator ground state of $A$, and $\psi_B,\psi_B'$ are arbitrary states in $B$.  However, $|\psi_{AB}\rangle$ only involves states
from system A with one fermion per site, and thus
\beq
\langle \psi_B'|H^B_{eff}|\psi_B\rangle = -g\langle \psi^s_A \otimes \psi_B'|\psi_{AB}\rangle\langle \psi_{AB}|\psi^s_A\otimes \psi_B\rangle, \label{effective}
\eeq
where $\psi^s_A$ is the Gutzwiller projection of the Chern insulator ground state $\psi^0_A$, which in turn  is a chiral spin liquid.

Now, because  $\psi_{AB}$ is a maximally entangled state\footnote{The reduced density matrix is the unit matrix.}, it can be expressed via  a Schmidt decomposition 
as
\beq
|\psi_{AB}\rangle = \frac{1}{\sqrt{2^N}}\sum_I |\alpha_I\rangle_A \otimes |{\tilde \alpha_I}\rangle_B, \label{schmidt}
\eeq
for {\it any} orthonormal basis
 $\big \{ |\alpha_I\rangle_A\big \}$
of the spin sector (one fermion per site) 
of system $A$.
 Here $|{\tilde \alpha_I}\rangle_B \equiv \sqrt{2^N} \  {}_A\langle\alpha_I|\psi_{AB}\rangle$ is the corresponding basis for system $B$. 
 For $|\psi_{AB}\rangle$, the product state of singlets, it is clear that when written in the basis
$|\alpha_I\rangle_A 
=
|\{\sigma_i\}\rangle_A 
$
 used in Eq. (\ref{LabelEqMaxEntangled}),
that
$|{\tilde \alpha_I}\rangle_B = \Theta |\alpha_I\rangle_A$
where $\Theta$ is the time-reversal operator.
But 
it is straightforward \cite{SM} to show that 
this statement 
holds in {\it any} basis $\big \{ |\alpha_I\rangle_A \big \}$ of the spin-sector (one fermion per site) of system A.
%
%
Thus, choosing a basis $\big \{ |\alpha_I\rangle_A \big \}$ 
of which $\psi^s_A$, the Gutzwiller projected ground state  of A  is an element, 
(\ref{effective}) and (\ref{schmidt}) yield
\beq
H^B_{eff} = -g |\overline{\psi^s_B}\rangle \langle \overline{\psi^s_B}| , \label{Bham}
\eeq
where $|\overline{\psi^s_B}\rangle$ is the time-reversed counterpart of the Gutzwiller projected Chern insulator ground state: 
it is a chiral spin liquid with 
chirality 
opposite to
the parent Chern insulator.

Although the original Kondo coupling ${\hat H}_K$  is a sum
of  {\it  local}  projectors onto a spin singlet (one at each lattice site $i$), we can interpolate between the exact limit of  the
{\it global}  projector employed in Eq. (\ref{LabelEqGlobalProjector}),
and the {\it local}  projector appearing  in the original Kondo interaction term ${\hat H}_K$ of Eq. (\ref{main}). 
In particular, we partition systems A and B into regions $r$ of linear dimension $l>\xi$, where $\xi$ is the correlation length of $A$, and consider the 
{\it quasi-local}  projector $P_r=-g |\psi_{AB,r}\rangle\langle \psi_{AB,r}|$ onto the product state of singlets\footnote{$|\psi_{AB,r}\rangle$
is analogous
to Eq. (\ref{LabelEqMaxEntangled}) but constrained to a region $r$ of linear size $\ell$.}
 supported in region $r$ (See Fig. 2). The sum over these
 {\it quasi-local} projectors, multiplied by $(-g)$,  now replaces  the {\it global} projector term  in 
Eq. (\ref{LabelEqGlobalProjector}).

\begin{figure}
\centering
\includegraphics[width=3.5in]{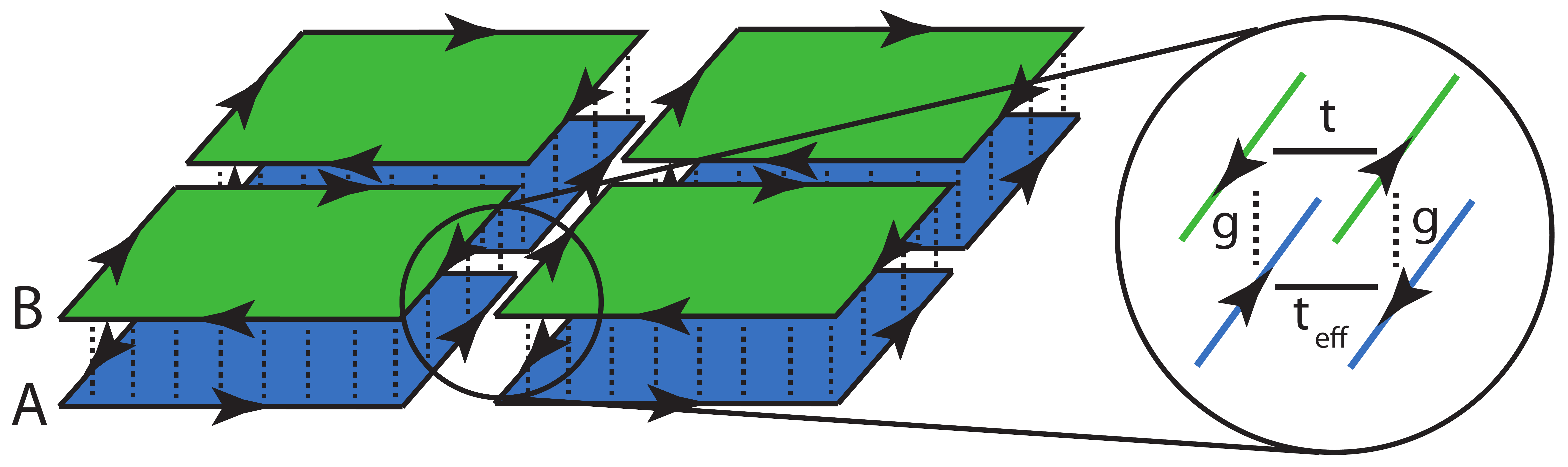}
\caption{Systems $A$ and $B$ are each partitioned into regions.  Corresponding regions $r$ in $A$ and $B$ are coupled via a projection $P_r$ onto the maximally entangled state of the two regions (dotted lines).  If adjacent $A$ regions are decoupled, as shown on the left, then each region of Chern insulator induces a region of chiral spin liquid via Eq. (\ref{Bham}).  Once the $A$ regions are recoupled, adjacent edge states gap out and the regions merge into a large Chern insulator.  Such interactions induce analogous interactions in $B$ (inset), which fuse the $B$ regions into a large chiral spin liquid. 
}
\end{figure}

First we consider turning off the interactions $H^A_{rr'}$ between adjacent regions $r,r'$ in $A$.  Each region of the Chern insulator
in A  gives rise to a corresponding region of chiral spin liquid in B  via the projector $P_r$, as shown in Fig. 2. The low energy states are the chiral (anti-chiral) edge modes of the two layers A and B, weakly coupled by an interlayer
 interaction of order $g$.  However, once the interaction $H^A_{rr'}$
between adjacent regions $r$ and $r'$ in region A is restored, the thereby generated
coupling between adjacent edge states within $A$ stitches the Chern insulator back together.  At the same time, $H^A_{rr'}$ induces via the interlayer coupling a corresponding interaction $H^B_{rr'}$ in layer B (see Fig. 2 inset) which gaps out adjacent edge states in $B$, thus creating a chiral spin liquid state in the whole of system B.

By choosing finer partitions, this construction interpolates between the {\it global  projector}, and {\it local projector} considered in the primary model.  Allowing for the full extent of interpolation requires as small a correlation length $\xi$ for system $A$ as possible,
since the linear size of the quasi-local projectors is limited by $\xi$.  While the  original model with a local Kondo coupling is an extreme limit of this interpolation, we note that the strength of the coupling $g$ serves as an additional parameter which can also interpolate between the global and local projectors; the larger $g$ is, the more both projectors favor the maximally entangled state.  Nevertheless, the validity of this picture for the local Kondo coupling model ultimately requires further analysis for specific lattice models, which we now present.

\begin{figure}
\centering
\includegraphics[width=3.5in]{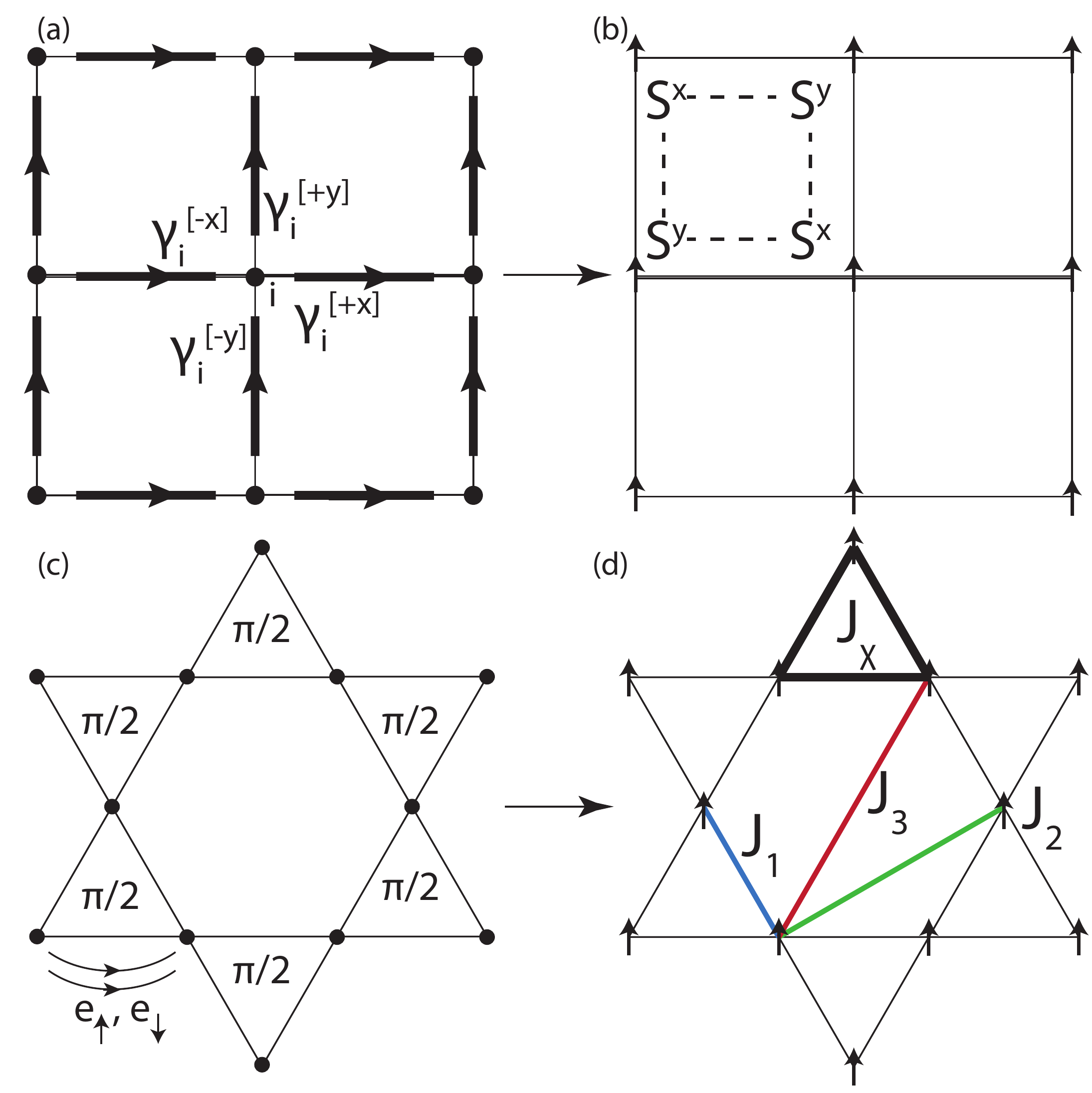}
\caption{Topological bootstrap in lattice models: (a) Free-fermion superconductor. Each site has four Majorana fermions, which form directed dimers (bold, with arrows) on the links with neighboring modes. (b) The superconductor induces the Wen plaquette/toric code via Kondo coupling. (c) Chern insulator on the kagome lattice realized from nearest neighbor hopping of spin up and down electrons with a background $\pi/2$ flux per triangle. (d) The Chern insulator induces a chiral spin liquid via Kondo coupling, with exchange and chirality parameters given in (\ref{parameters}).}
\label{LabelFigThreeTopBootstrapLatticeModels}
\end{figure}

\section{$Z_2$ Spin Liquids from Superconductors}

We first present
an exactly solvable example for the entanglement argument provided in the previous section. In particular, we consider
an example in which the parent system (layer A) is a free-fermion superconductor with zero correlation length, 
from which we can show analytically that the 
topological bootstrap 
(with local Kondo coupling) gives rise to a $Z_2$ spin liquid -- the Gutzwiller projection of the free-fermion superconductor, predicted
by the entanglement argument of the previous section. The effective Hamiltonian induced in $B$ is exactly the toric code\cite{Kitaev2006,Wen2003}.

Let 
system $A$  be a square lattice with four Majorana modes per site $i$, labeled $\gamma_i^{[\pm x]}, \gamma_i^{[\pm y]}$ (see Fig. 3),
where
$\{ \gamma^{(\mu)}_i ,\gamma^{(\nu)}_j \} =2\delta_{i,j} \delta_{\mu,\nu}$.
System B 
consists of  spin-$1/2$ degrees of freedom  $S_i$ at each site $i$ of the square lattice. 
We choose an arrow convention on the links of the square lattice so that arrows are always directed along the $+x(y)$ direction for horizontal (vertical) links.
(See Fig. \ref{LabelFigThreeTopBootstrapLatticeModels}.)

We consider the following Hamiltonian
\beq
H&=&H_A+H_{AB} \\
H_A &=& 
\sum_i \sum_{\mu=+x,+y} i\gamma_i^{[+\mu]} \gamma_{i+\mu}^{[-\mu]} \\
H_{AB} &=&{g}\sum_{i,\alpha,\beta} {\vec S}_i\cdot \left ( c_{i,\alpha}^\dagger\vec\sigma_{\alpha\beta} c_{i,\beta}\right ) \\
c_{i,\uparrow}&=&\frac{\gamma^{[-x]}_i+\imth\gamma^{[+x]}_i}2,~~c_{i,\downarrow}=\frac{\gamma^{[+y]}_i+\imth\gamma^{[-y]}_i}2.
\eeq
The ground state of $H_A$ features dimerized Majorana fermions on every link (Fig. 3): this is a free-fermion topological superconductor with gapless Majorana edge states protected by translation symmetry (cutting through a rung of dimers exposes unpaired Majorana modes, which must be gapless given translation symmetry along the cut \cite{susy}). 
The term $H_{AB}$ 
represents an  exchange coupling with the initially decoupled spin-1/2 degrees of freedom of layer B.


At lowest order in $g$ in degenerate perturbation theory, one finds that the induced spin Hamiltonian in layer B  is
\beq
H^{eff}_B \propto g^4 \sum_i S^y_i S^x_{i+x} S^y_{i+x+y} S^x_{i+y},
\eeq
which is the Wen plaquette model\cite{Wen2003} with $Z_2$ topological order, equivalent to the toric code \cite{Kitaev2006} by a basis change.  As discussed in \cite{Wen2003}, the ground state of this model is the projection of the Majorana dimer wavefunction onto the sector with $ \gamma^{[+y]}_i \gamma^{[+x]}_i \gamma^{[-x]}_i \gamma^{[-y]}_i = 1$ for every site $i$, which corresponds to the
single-occupancy constraint $\sum_{\sigma} c^\dagger_{i,\sigma}c_{i,\sigma}=1$.  This solvable example with zero correlation length is a direct illustration of the entanglement argument provided
in the previous section.

\section{Chiral Spin Liquids from Chern Insulators}

We expect the above physics to persist when the correlation length of the $A$ system is small but finite, and we confirm this expectation with two lattice models.  First, we consider a model on the kagome lattice of the form (\ref{main}) and explicitly derive the
 spin Hamiltonian induced by the Chern insulator
to leading order in the Kondo  coupling $g$.

We use a tight-binding model\cite{Hermele2008} of fermions with nearest neighbor hopping in a background flux of $\phi$ through each triangle and $\pi-2\phi$ through each hexagon (see Fig.3).  
The hopping amplitude is set to $1$.
When $\phi=0,\pi$, the band structure is gapless with two Dirac cones at low energy. By taking $\phi=\pi/2$ a nontrivial gap is opened with a Chern number $1$ ground state.

The effective spin Hamiltonian is constrained by $SU(2)$ spin rotation and lattice symmetries to be of the form
\beq
H_{eff}^B = \sum_{ij} J_{ij} S_i \cdot S_j + \sum_{ijk} J_{ijk} S_i \cdot (S_j \times S_k) + ..., \label{spinham}
\eeq
where higher order spin interactions have been neglected because they are suppressed by powers of $g$.  Furthermore, because the fermionic system in layer A is gapped, the induced spin interactions in layer B  will decay exponentially with distance, and we can restrict attention to nearest neighbors for both the two-spin Heisenberg interaction and the three-spin ``chirality'' term, which requires broken time reversal symmetry.  Following the convention of previously considered kagome spin models, we will 
denote  the relevant two-spin couplings by $J_1,J_2, J_3$
(nearest neighbor, next nearest neighbor, 3rd nearest neighbor) and the triangular spin chirality term by $J_{\chi}$ (Fig. 3d).

These effective couplings can be attained from degenerate perturbation theory in $g$ 
(see Supplementary Material \cite{SM}), and for the above model 
we find,
after normalizing $J_1$ to 1:
\beq
J_1=1, J_2=0.51, J_3=0.56, J_{\chi}=-0.49g \label{parameters}
\eeq
First we consider $0 < g<<1$, for which $J_{\chi}$ can be ignored:
 A DMRG study\cite{Gong2014} established the chiral spin liquid phase in the parameter range $0.1J_1 < J_2=J_3<0.7J_1$, and a subsequent variational Monte Carlo study\cite{Hu2015} revealed that the CSL phase persists for $J_3>J_2$. Hence, the effective spin Hamiltonian from our model lies within the CSL phase.

For larger $g$, the next leading order contribution is the chirality term $J_{\chi}$.  The DMRG analysis in \cite{Bauer2014} has shown that the spin model with $J_{\chi}$ alone is a CSL, and that this phase persists even after including a nearest neighbor term as large as $J_1\approx 6J_{\chi}$. Thus, the CSL phase is reinforced by the higher order chirality term.

The topological induction of the kagome chiral spin liquid serves as a demonstration of concept for our construction, and we can apply it to many different lattices.  For example, consider a similar setup on the triangular lattice, in which there is also a simple tight-binding Hamiltonian realizing a Chern insulator with Chern number $1$: nearest neighbor hopping with a uniform $\pi/2$ flux through each triangle.  We use the same technique as above to derive the effective spin Hamiltonian (\ref{spinham}) that is induced, and we find at lowest order in $g$:
\beq
J_1=1, J_2=0.064, J_3=-0.018, J_{\chi}=-0.25g \nonumber
\eeq
With $J_1, J_2$ alone, a DMRG study\cite{Zhu2015} found evidence for a gapped spin liquid in the regime $0.06<J_2/J_1<0.17$ while another DMRG study\cite{Hu2015a} has found spin liquids in the regime $0.07<J_2/J_1<0.15$, with evidence of broken time-reversal symmetry. To our knowledge, the nature of the spin liquid(s) in this regime has not been fully settled.  A more recent DMRG study finds that an additional spin chirality term $J_\chi$ as small as $0.02J_1$ can stabilize the chiral spin liquid, and a variational Monte Carlo study\cite{Hu2016} also advocates the role of the chirality term in stabilizing the CSL.  Finally, a recent exact diagonalization study\cite{Wietek2016} also indicates that given the above $J_2$, a small chirality term favors the CSL.  These numerical studies indicate that our model at small but finite $g$ is in the chiral spin liquid phase.

\section{Summary and Discussion}

We have provided a powerful machinery to achieve intrinsic topological orders, such as the $Z_2$ and chiral spin liquids, from accessible free fermion phases, such as superconductors and Chern insulators. General arguments involving the topological proximity effect and entanglement, as well as explicit lattice models treated analytically and numerically, support this induction of intrinsic topological orders via Kondo coupling to a free fermion phase.  Two elements are essential for this topological bootstrap.  On one hand, the bulk topological proximity effect provides a mechanism of inducing the inverse topological phase of the parent system.  On the other hand, the fact that the secondary system only hosts spin degrees of freedom forces the induction to be restricted to the spin sector of the parent system, hence realizing the Gutzwiller-projected free fermion states as the ground state of the spin system.

The maximally entangled state of local singlets thus serves as a channel through which the spin sector of a non-interacting system of fermions can be inverted and filtered into the localized spin system. There are many directions to explore this topological bootstrap scheme, including different lattices, higher dimensions, different parent free fermion phases, and alternative coupling schemes.  Our construction provides both a physical route toward realizing 
phases with intrinsic topological order
 in synthetic composite systems such as ultracold atoms as well as a numerical route for deriving microscopic spin Hamiltonians which realize quantum spin liquids. As the Gutzwiller projection of higher Chern number states results in non-abelian chiral spin liquids\cite{Wen1999}, it would be interesting to use the topological bootstrap to generate lattice models of such non-abelian phases.

{\it Acknowledgements}: We thank S-S. Gong, T. Grover, W-J. Hu, and H-H. Lai for useful discussions. THH is supported by a fellowship from the Gordon and Betty Moore Foundation (Grant 4304). YML is supported by startup funds at Ohio State University.  AWWL is supported by NSF under Grant No. DMR-1309667.

\bibliographystyle{apsrev4-1}
\bibliography{bibs}

\begin{thebibliography}{32}%
\makeatletter
\providecommand \@ifxundefined [1]{%
 \@ifx{#1\undefined}
}%
\providecommand \@ifnum [1]{%
 \ifnum #1\expandafter \@firstoftwo
 \else \expandafter \@secondoftwo
 \fi
}%
\providecommand \@ifx [1]{%
 \ifx #1\expandafter \@firstoftwo
 \else \expandafter \@secondoftwo
 \fi
}%
\providecommand \natexlab [1]{#1}%
\providecommand \enquote  [1]{``#1''}%
\providecommand \bibnamefont  [1]{#1}%
\providecommand \bibfnamefont [1]{#1}%
\providecommand \citenamefont [1]{#1}%
\providecommand \href@noop [0]{\@secondoftwo}%
\providecommand \href [0]{\begingroup \@sanitize@url \@href}%
\providecommand \@href[1]{\@@startlink{#1}\@@href}%
\providecommand \@@href[1]{\endgroup#1\@@endlink}%
\providecommand \@sanitize@url [0]{\catcode `\\12\catcode `\$12\catcode
  `\&12\catcode `\#12\catcode `\^12\catcode `\_12\catcode `\%12\relax}%
\providecommand \@@startlink[1]{}%
\providecommand \@@endlink[0]{}%
\providecommand \url  [0]{\begingroup\@sanitize@url \@url }%
\providecommand \@url [1]{\endgroup\@href {#1}{\urlprefix }}%
\providecommand \urlprefix  [0]{URL }%
\providecommand \Eprint [0]{\href }%
\providecommand \doibase [0]{http://dx.doi.org/}%
\providecommand \selectlanguage [0]{\@gobble}%
\providecommand \bibinfo  [0]{\@secondoftwo}%
\providecommand \bibfield  [0]{\@secondoftwo}%
\providecommand \translation [1]{[#1]}%
\providecommand \BibitemOpen [0]{}%
\providecommand \bibitemStop [0]{}%
\providecommand \bibitemNoStop [0]{.\EOS\space}%
\providecommand \EOS [0]{\spacefactor3000\relax}%
\providecommand \BibitemShut  [1]{\csname bibitem#1\endcsname}%
\let\auto@bib@innerbib\@empty
\bibitem [{\citenamefont {Chen}\ \emph {et~al.}(2013)\citenamefont {Chen},
  \citenamefont {Gu}, \citenamefont {Liu},\ and\ \citenamefont
  {Wen}}]{chen2013}%
  \BibitemOpen
  \bibfield  {author} {\bibinfo {author} {\bibfnamefont {X.}~\bibnamefont
  {Chen}}, \bibinfo {author} {\bibfnamefont {Z.-C.}\ \bibnamefont {Gu}},
  \bibinfo {author} {\bibfnamefont {Z.-X.}\ \bibnamefont {Liu}}, \ and\
  \bibinfo {author} {\bibfnamefont {X.-G.}\ \bibnamefont {Wen}},\ }\href@noop
  {} {\bibfield  {journal} {\bibinfo  {journal} {PRB 87, 155114}\ } (\bibinfo
  {year} {2013})}\BibitemShut {NoStop}%
\bibitem [{\citenamefont {Pollmann}\ \emph {et~al.}(2010)\citenamefont
  {Pollmann}, \citenamefont {Turner}, \citenamefont {Berg},\ and\ \citenamefont
  {Oshikawa}}]{pollmann2010}%
  \BibitemOpen
  \bibfield  {author} {\bibinfo {author} {\bibfnamefont {F.}~\bibnamefont
  {Pollmann}}, \bibinfo {author} {\bibfnamefont {A.}~\bibnamefont {Turner}},
  \bibinfo {author} {\bibfnamefont {E.}~\bibnamefont {Berg}}, \ and\ \bibinfo
  {author} {\bibfnamefont {M.}~\bibnamefont {Oshikawa}},\ }\href@noop {}
  {\bibfield  {journal} {\bibinfo  {journal} {Phys. Rev. B 81, 064439}\ }
  (\bibinfo {year} {2010})}\BibitemShut {NoStop}%
\bibitem [{\citenamefont {Gu}\ and\ \citenamefont {Wen}(2009)}]{wen2009}%
  \BibitemOpen
  \bibfield  {author} {\bibinfo {author} {\bibfnamefont {Z.-C.}\ \bibnamefont
  {Gu}}\ and\ \bibinfo {author} {\bibfnamefont {X.-G.}\ \bibnamefont {Wen}},\
  }\href@noop {} {\bibfield  {journal} {\bibinfo  {journal} {PRB 80, 155131}\ }
  (\bibinfo {year} {2009})}\BibitemShut {NoStop}%
\bibitem [{\citenamefont {Wen}(1990)}]{wen1990}%
  \BibitemOpen
  \bibfield  {author} {\bibinfo {author} {\bibfnamefont {X.-G.}\ \bibnamefont
  {Wen}},\ }\href@noop {} {\bibfield  {journal} {\bibinfo  {journal} {Int. J.
  Mod. Phys. B4, 239}\ } (\bibinfo {year} {1990})}\BibitemShut {NoStop}%
\bibitem [{\citenamefont {Hasan}\ and\ \citenamefont {Kane}(2010)}]{hasan2010}%
  \BibitemOpen
  \bibfield  {author} {\bibinfo {author} {\bibfnamefont {M.}~\bibnamefont
  {Hasan}}\ and\ \bibinfo {author} {\bibfnamefont {C.}~\bibnamefont {Kane}},\
  }\href@noop {} {\bibfield  {journal} {\bibinfo  {journal} {Rev. Mod. Phys.
  82, 3045}\ } (\bibinfo {year} {2010})}\BibitemShut {NoStop}%
\bibitem [{\citenamefont {Fu}(2011)}]{fu2011}%
  \BibitemOpen
  \bibfield  {author} {\bibinfo {author} {\bibfnamefont {L.}~\bibnamefont
  {Fu}},\ }\href@noop {} {\bibfield  {journal} {\bibinfo  {journal} {Phys. Rev.
  Lett. 106, 106802}\ } (\bibinfo {year} {2011})}\BibitemShut {NoStop}%
\bibitem [{\citenamefont {Hsieh}\ \emph {et~al.}(2012)\citenamefont {Hsieh},
  \citenamefont {Lin}, \citenamefont {Liu}, \citenamefont {Duan}, \citenamefont
  {Bansil},\ and\ \citenamefont {Fu}}]{hsieh2012}%
  \BibitemOpen
  \bibfield  {author} {\bibinfo {author} {\bibfnamefont {T.}~\bibnamefont
  {Hsieh}}, \bibinfo {author} {\bibfnamefont {H.}~\bibnamefont {Lin}}, \bibinfo
  {author} {\bibfnamefont {J.}~\bibnamefont {Liu}}, \bibinfo {author}
  {\bibfnamefont {W.}~\bibnamefont {Duan}}, \bibinfo {author} {\bibfnamefont
  {A.}~\bibnamefont {Bansil}}, \ and\ \bibinfo {author} {\bibfnamefont
  {L.}~\bibnamefont {Fu}},\ }\href@noop {} {\bibfield  {journal} {\bibinfo
  {journal} {Nat. Comm. 3, 982}\ } (\bibinfo {year} {2012})}\BibitemShut
  {NoStop}%
\bibitem [{\citenamefont {Tsui}\ \emph {et~al.}(1982)\citenamefont {Tsui},
  \citenamefont {Stormer},\ and\ \citenamefont {Gossard}}]{Tsui1982}%
  \BibitemOpen
  \bibfield  {author} {\bibinfo {author} {\bibfnamefont {D.}~\bibnamefont
  {Tsui}}, \bibinfo {author} {\bibfnamefont {H.}~\bibnamefont {Stormer}}, \
  and\ \bibinfo {author} {\bibfnamefont {A.}~\bibnamefont {Gossard}},\
  }\href@noop {} {\bibfield  {journal} {\bibinfo  {journal} {Phys. Rev. Lett.
  48 (22) 1559}\ } (\bibinfo {year} {1982})}\BibitemShut {NoStop}%
\bibitem [{\citenamefont {Balents}(2010)}]{balents2010}%
  \BibitemOpen
  \bibfield  {author} {\bibinfo {author} {\bibfnamefont {L.}~\bibnamefont
  {Balents}},\ }\href@noop {} {\bibfield  {journal} {\bibinfo  {journal}
  {Nature 464, 199}\ } (\bibinfo {year} {2010})}\BibitemShut {NoStop}%
\bibitem [{\citenamefont {Nayak}\ \emph {et~al.}(2008)\citenamefont {Nayak},
  \citenamefont {Simon}, \citenamefont {Stern}, \citenamefont {Freedman},\ and\
  \citenamefont {Sarma}}]{nayak2008}%
  \BibitemOpen
  \bibfield  {author} {\bibinfo {author} {\bibfnamefont {C.}~\bibnamefont
  {Nayak}}, \bibinfo {author} {\bibfnamefont {S.}~\bibnamefont {Simon}},
  \bibinfo {author} {\bibfnamefont {A.}~\bibnamefont {Stern}}, \bibinfo
  {author} {\bibfnamefont {M.}~\bibnamefont {Freedman}}, \ and\ \bibinfo
  {author} {\bibfnamefont {S.}~\bibnamefont {Sarma}},\ }\href@noop {}
  {\bibfield  {journal} {\bibinfo  {journal} {Rev. Mod. Phys. 80, 1083}\ }
  (\bibinfo {year} {2008})}\BibitemShut {NoStop}%
\bibitem [{\citenamefont {Kitaev}(2003)}]{kitaev2003}%
  \BibitemOpen
  \bibfield  {author} {\bibinfo {author} {\bibfnamefont {A.}~\bibnamefont
  {Kitaev}},\ }\href@noop {} {\bibfield  {journal} {\bibinfo  {journal} {Ann.
  Phys. N.Y. 303, 2}\ } (\bibinfo {year} {2003})}\BibitemShut {NoStop}%
\bibitem [{\citenamefont {Anderson}(1987)}]{Anderson1987}%
  \BibitemOpen
  \bibfield  {author} {\bibinfo {author} {\bibfnamefont {P.~W.}\ \bibnamefont
  {Anderson}},\ }\href
  {http://science.sciencemag.org/content/235/4793/1196.abstract} {\bibfield
  {journal} {\bibinfo  {journal} {Science}\ }\textbf {\bibinfo {volume}
  {235}},\ \bibinfo {pages} {1196} (\bibinfo {year} {1987})}\BibitemShut
  {NoStop}%
\bibitem [{\citenamefont {Gros}(1989)}]{Gros1989}%
  \BibitemOpen
  \bibfield  {author} {\bibinfo {author} {\bibfnamefont {C.}~\bibnamefont
  {Gros}},\ }\href
  {http://www.sciencedirect.com/science/article/pii/0003491689900778}
  {\bibfield  {journal} {\bibinfo  {journal} {Annals of Physics}\ }\textbf
  {\bibinfo {volume} {189}},\ \bibinfo {pages} {53} (\bibinfo {year}
  {1989})}\BibitemShut {NoStop}%
\bibitem [{\citenamefont {Wen}(1999)}]{Wen1999}%
  \BibitemOpen
  \bibfield  {author} {\bibinfo {author} {\bibfnamefont {X.-G.}\ \bibnamefont
  {Wen}},\ }\href {http://link.aps.org/doi/10.1103/PhysRevB.60.8827} {\bibfield
   {journal} {\bibinfo  {journal} {Phys. Rev. B}\ }\textbf {\bibinfo {volume}
  {60}},\ \bibinfo {pages} {8827} (\bibinfo {year} {1999})}\BibitemShut
  {NoStop}%
\bibitem [{\citenamefont {Wen}(2002)}]{Wen2002}%
  \BibitemOpen
  \bibfield  {author} {\bibinfo {author} {\bibfnamefont {X.-G.}\ \bibnamefont
  {Wen}},\ }\href {http://link.aps.org/doi/10.1103/PhysRevB.65.165113}
  {\bibfield  {journal} {\bibinfo  {journal} {Phys. Rev. B}\ }\textbf {\bibinfo
  {volume} {65}},\ \bibinfo {pages} {165113} (\bibinfo {year}
  {2002})}\BibitemShut {NoStop}%
\bibitem [{\citenamefont {Lee}\ \emph {et~al.}(2006)\citenamefont {Lee},
  \citenamefont {Nagaosa},\ and\ \citenamefont {Wen}}]{Lee2006}%
  \BibitemOpen
  \bibfield  {author} {\bibinfo {author} {\bibfnamefont {P.~A.}\ \bibnamefont
  {Lee}}, \bibinfo {author} {\bibfnamefont {N.}~\bibnamefont {Nagaosa}}, \ and\
  \bibinfo {author} {\bibfnamefont {X.-G.}\ \bibnamefont {Wen}},\ }\href
  {http://link.aps.org/doi/10.1103/RevModPhys.78.17} {\bibfield  {journal}
  {\bibinfo  {journal} {Rev. Mod. Phys.}\ }\textbf {\bibinfo {volume} {78}},\
  \bibinfo {pages} {17} (\bibinfo {year} {2006})}\BibitemShut {NoStop}%
\bibitem [{\citenamefont {Hsieh}\ \emph
  {et~al.}(2016{\natexlab{a}})\citenamefont {Hsieh}, \citenamefont {Ishizuka},
  \citenamefont {Balents},\ and\ \citenamefont {Hughes}}]{Hsieh2016}%
  \BibitemOpen
  \bibfield  {author} {\bibinfo {author} {\bibfnamefont {T.~H.}\ \bibnamefont
  {Hsieh}}, \bibinfo {author} {\bibfnamefont {H.}~\bibnamefont {Ishizuka}},
  \bibinfo {author} {\bibfnamefont {L.}~\bibnamefont {Balents}}, \ and\
  \bibinfo {author} {\bibfnamefont {T.~L.}\ \bibnamefont {Hughes}},\ }\href
  {http://link.aps.org/doi/10.1103/PhysRevLett.116.086802} {\bibfield
  {journal} {\bibinfo  {journal} {Phys. Rev. Lett.}\ }\textbf {\bibinfo
  {volume} {116}},\ \bibinfo {pages} {086802} (\bibinfo {year}
  {2016}{\natexlab{a}})}\BibitemShut {NoStop}%
\bibitem [{\citenamefont {Kalmeyer}\ and\ \citenamefont
  {Laughlin}(1987)}]{Kalmeyer1987}%
  \BibitemOpen
  \bibfield  {author} {\bibinfo {author} {\bibfnamefont {V.}~\bibnamefont
  {Kalmeyer}}\ and\ \bibinfo {author} {\bibfnamefont {R.~B.}\ \bibnamefont
  {Laughlin}},\ }\href {http://link.aps.org/doi/10.1103/PhysRevLett.59.2095}
  {\bibfield  {journal} {\bibinfo  {journal} {Phys. Rev. Lett.}\ }\textbf
  {\bibinfo {volume} {59}},\ \bibinfo {pages} {2095} (\bibinfo {year}
  {1987})}\BibitemShut {NoStop}%
\bibitem [{\citenamefont {Wen}\ \emph {et~al.}(1989)\citenamefont {Wen},
  \citenamefont {Wilczek},\ and\ \citenamefont {Zee}}]{Wen1989}%
  \BibitemOpen
  \bibfield  {author} {\bibinfo {author} {\bibfnamefont {X.~G.}\ \bibnamefont
  {Wen}}, \bibinfo {author} {\bibfnamefont {F.}~\bibnamefont {Wilczek}}, \ and\
  \bibinfo {author} {\bibfnamefont {A.}~\bibnamefont {Zee}},\ }\href
  {http://link.aps.org/doi/10.1103/PhysRevB.39.11413} {\bibfield  {journal}
  {\bibinfo  {journal} {Phys. Rev. B}\ }\textbf {\bibinfo {volume} {39}},\
  \bibinfo {pages} {11413} (\bibinfo {year} {1989})}\BibitemShut {NoStop}%
\bibitem [{\citenamefont {Nielsen}\ \emph {et~al.}(2013)\citenamefont
  {Nielsen}, \citenamefont {Sierra},\ and\ \citenamefont
  {Cirac}}]{Nielsen2013}%
  \BibitemOpen
  \bibfield  {author} {\bibinfo {author} {\bibfnamefont {A.~E.~B.}\
  \bibnamefont {Nielsen}}, \bibinfo {author} {\bibfnamefont {G.}~\bibnamefont
  {Sierra}}, \ and\ \bibinfo {author} {\bibfnamefont {J.~I.}\ \bibnamefont
  {Cirac}},\ }\href {http://dx.doi.org/10.1038/ncomms3864} {\bibfield
  {journal} {\bibinfo  {journal} {Nature Communications}\ }\textbf {\bibinfo
  {volume} {4}},\ \bibinfo {pages} {2864} (\bibinfo {year} {2013})}\BibitemShut
  {NoStop}%
\bibitem [{SM()}]{SM}%
  \BibitemOpen
  \href@noop {} {\bibinfo  {journal} {See Supplementary Material for a
  derivation of the time-reversed Gutzwiller projection and the derivation of
  the effective spin Hamiltonian via degenerate perturbation theory}\
  }\BibitemShut {NoStop}%
\bibitem [{\citenamefont {Kitaev}(2006)}]{Kitaev2006}%
  \BibitemOpen
\bibfield  {journal} {  }\bibfield  {author} {\bibinfo {author} {\bibfnamefont
  {A.}~\bibnamefont {Kitaev}},\ }\href@noop {} {\bibfield  {journal} {\bibinfo
  {journal} {Ann. Phys. 321, 2}\ } (\bibinfo {year} {2006})}\BibitemShut
  {NoStop}%
\bibitem [{\citenamefont {Wen}(2003)}]{Wen2003}%
  \BibitemOpen
  \bibfield  {author} {\bibinfo {author} {\bibfnamefont {X.-G.}\ \bibnamefont
  {Wen}},\ }\href {http://link.aps.org/doi/10.1103/PhysRevLett.90.016803}
  {\bibfield  {journal} {\bibinfo  {journal} {Phys. Rev. Lett.}\ }\textbf
  {\bibinfo {volume} {90}},\ \bibinfo {pages} {016803} (\bibinfo {year}
  {2003})}\BibitemShut {NoStop}%
\bibitem [{\citenamefont {Hsieh}\ \emph
  {et~al.}(2016{\natexlab{b}})\citenamefont {Hsieh}, \citenamefont {Hal\'asz},\
  and\ \citenamefont {Grover}}]{susy}%
  \BibitemOpen
  \bibfield  {author} {\bibinfo {author} {\bibfnamefont {T.}~\bibnamefont
  {Hsieh}}, \bibinfo {author} {\bibfnamefont {G.}~\bibnamefont {Hal\'asz}}, \
  and\ \bibinfo {author} {\bibfnamefont {T.}~\bibnamefont {Grover}},\
  }\href@noop {} {\bibfield  {journal} {\bibinfo  {journal} {Phys. Rev. Lett.
  117, 166802}\ } (\bibinfo {year} {2016}{\natexlab{b}})}\BibitemShut {NoStop}%
\bibitem [{\citenamefont {Hermele}\ \emph {et~al.}(2008)\citenamefont
  {Hermele}, \citenamefont {Ran}, \citenamefont {Lee},\ and\ \citenamefont
  {Wen}}]{Hermele2008}%
  \BibitemOpen
  \bibfield  {author} {\bibinfo {author} {\bibfnamefont {M.}~\bibnamefont
  {Hermele}}, \bibinfo {author} {\bibfnamefont {Y.}~\bibnamefont {Ran}},
  \bibinfo {author} {\bibfnamefont {P.~A.}\ \bibnamefont {Lee}}, \ and\
  \bibinfo {author} {\bibfnamefont {X.-G.}\ \bibnamefont {Wen}},\ }\href
  {http://link.aps.org/doi/10.1103/PhysRevB.77.224413} {\bibfield  {journal}
  {\bibinfo  {journal} {Phys. Rev. B}\ }\textbf {\bibinfo {volume} {77}},\
  \bibinfo {pages} {224413} (\bibinfo {year} {2008})}\BibitemShut {NoStop}%
\bibitem [{\citenamefont {Gong}\ \emph {et~al.}(2014)\citenamefont {Gong},
  \citenamefont {Zhu},\ and\ \citenamefont {Sheng}}]{Gong2014}%
  \BibitemOpen
  \bibfield  {author} {\bibinfo {author} {\bibfnamefont {S.-S.}\ \bibnamefont
  {Gong}}, \bibinfo {author} {\bibfnamefont {W.}~\bibnamefont {Zhu}}, \ and\
  \bibinfo {author} {\bibfnamefont {D.~N.}\ \bibnamefont {Sheng}},\ }\href
  {http://dx.doi.org/10.1038/srep06317} {\bibfield  {journal} {\bibinfo
  {journal} {Scientific Reports}\ }\textbf {\bibinfo {volume} {4}},\ \bibinfo
  {pages} {6317} (\bibinfo {year} {2014})}\BibitemShut {NoStop}%
\bibitem [{\citenamefont {Hu}\ \emph {et~al.}(2015{\natexlab{a}})\citenamefont
  {Hu}, \citenamefont {Zhu}, \citenamefont {Zhang}, \citenamefont {Gong},
  \citenamefont {Becca},\ and\ \citenamefont {Sheng}}]{Hu2015}%
  \BibitemOpen
  \bibfield  {author} {\bibinfo {author} {\bibfnamefont {W.-J.}\ \bibnamefont
  {Hu}}, \bibinfo {author} {\bibfnamefont {W.}~\bibnamefont {Zhu}}, \bibinfo
  {author} {\bibfnamefont {Y.}~\bibnamefont {Zhang}}, \bibinfo {author}
  {\bibfnamefont {S.}~\bibnamefont {Gong}}, \bibinfo {author} {\bibfnamefont
  {F.}~\bibnamefont {Becca}}, \ and\ \bibinfo {author} {\bibfnamefont {D.~N.}\
  \bibnamefont {Sheng}},\ }\href
  {http://link.aps.org/doi/10.1103/PhysRevB.91.041124} {\bibfield  {journal}
  {\bibinfo  {journal} {Phys. Rev. B}\ }\textbf {\bibinfo {volume} {91}},\
  \bibinfo {pages} {041124} (\bibinfo {year} {2015}{\natexlab{a}})}\BibitemShut
  {NoStop}%
\bibitem [{\citenamefont {Bauer}\ \emph {et~al.}(2014)\citenamefont {Bauer},
  \citenamefont {Cincio}, \citenamefont {Keller}, \citenamefont {Dolfi},
  \citenamefont {Vidal}, \citenamefont {Trebst},\ and\ \citenamefont
  {Ludwig}}]{Bauer2014}%
  \BibitemOpen
  \bibfield  {author} {\bibinfo {author} {\bibfnamefont {B.}~\bibnamefont
  {Bauer}}, \bibinfo {author} {\bibfnamefont {L.}~\bibnamefont {Cincio}},
  \bibinfo {author} {\bibfnamefont {B.~P.}\ \bibnamefont {Keller}}, \bibinfo
  {author} {\bibfnamefont {M.}~\bibnamefont {Dolfi}}, \bibinfo {author}
  {\bibfnamefont {G.}~\bibnamefont {Vidal}}, \bibinfo {author} {\bibfnamefont
  {S.}~\bibnamefont {Trebst}}, \ and\ \bibinfo {author} {\bibfnamefont
  {A.~W.~W.}\ \bibnamefont {Ludwig}},\ }\href
  {http://dx.doi.org/10.1038/ncomms6137} {\bibfield  {journal} {\bibinfo
  {journal} {Nature Communications}\ }\textbf {\bibinfo {volume} {5}},\
  \bibinfo {pages} {5137} (\bibinfo {year} {2014})}\BibitemShut {NoStop}%
\bibitem [{\citenamefont {Zhu}\ and\ \citenamefont {White}(2015)}]{Zhu2015}%
  \BibitemOpen
  \bibfield  {author} {\bibinfo {author} {\bibfnamefont {Z.}~\bibnamefont
  {Zhu}}\ and\ \bibinfo {author} {\bibfnamefont {S.~R.}\ \bibnamefont
  {White}},\ }\href {http://link.aps.org/doi/10.1103/PhysRevB.92.041105}
  {\bibfield  {journal} {\bibinfo  {journal} {Phys. Rev. B}\ }\textbf {\bibinfo
  {volume} {92}},\ \bibinfo {pages} {041105} (\bibinfo {year}
  {2015})}\BibitemShut {NoStop}%
\bibitem [{\citenamefont {Hu}\ \emph {et~al.}(2015{\natexlab{b}})\citenamefont
  {Hu}, \citenamefont {Gong}, \citenamefont {Zhu},\ and\ \citenamefont
  {Sheng}}]{Hu2015a}%
  \BibitemOpen
  \bibfield  {author} {\bibinfo {author} {\bibfnamefont {W.-J.}\ \bibnamefont
  {Hu}}, \bibinfo {author} {\bibfnamefont {S.-S.}\ \bibnamefont {Gong}},
  \bibinfo {author} {\bibfnamefont {W.}~\bibnamefont {Zhu}}, \ and\ \bibinfo
  {author} {\bibfnamefont {D.~N.}\ \bibnamefont {Sheng}},\ }\href
  {http://link.aps.org/doi/10.1103/PhysRevB.92.140403} {\bibfield  {journal}
  {\bibinfo  {journal} {Phys. Rev. B}\ }\textbf {\bibinfo {volume} {92}},\
  \bibinfo {pages} {140403} (\bibinfo {year} {2015}{\natexlab{b}})}\BibitemShut
  {NoStop}%
\bibitem [{\citenamefont {Hu}\ \emph {et~al.}(2016)\citenamefont {Hu},
  \citenamefont {Gong},\ and\ \citenamefont {Sheng}}]{Hu2016}%
  \BibitemOpen
  \bibfield  {author} {\bibinfo {author} {\bibfnamefont {W.-J.}\ \bibnamefont
  {Hu}}, \bibinfo {author} {\bibfnamefont {S.-S.}\ \bibnamefont {Gong}}, \ and\
  \bibinfo {author} {\bibfnamefont {D.~N.}\ \bibnamefont {Sheng}},\ }\href
  {http://link.aps.org/doi/10.1103/PhysRevB.94.075131} {\bibfield  {journal}
  {\bibinfo  {journal} {Phys. Rev. B}\ }\textbf {\bibinfo {volume} {94}},\
  \bibinfo {pages} {075131} (\bibinfo {year} {2016})}\BibitemShut {NoStop}%
\bibitem [{\citenamefont {{Wietek}}\ and\ \citenamefont
  {{Lauchli}}(2016)}]{Wietek2016}%
  \BibitemOpen
  \bibfield  {author} {\bibinfo {author} {\bibfnamefont {A.}~\bibnamefont
  {{Wietek}}}\ and\ \bibinfo {author} {\bibfnamefont {A.~M.}\ \bibnamefont
  {{Lauchli}}},\ }\href {https://arxiv.org/abs/1604.07829} {\bibfield
  {journal} {\bibinfo  {journal} {ArXiv e-prints}\ } (\bibinfo {year}
  {2016})},\ \Eprint {http://arxiv.org/abs/1604.07829} {arXiv:1604.07829
  [cond-mat.str-el]} \BibitemShut {NoStop}%
\end{thebibliography}%

\end{document}